\documentclass[aps,pra,floatfix,noshowkeys,epsfig,graphics,natbib]{revtex4}
\usepackage{graphicx}
\usepackage{amsmath}
\usepackage{amsfonts}
\usepackage{amssymb}
\usepackage{comment}
\usepackage{xcolor}
\usepackage{kotex}
\usepackage{natbib}
\usepackage{lineno}

\newcommand{\newc}{\newcommand}
\newc{\beq}    {\begin{equation}}
\newc{\eeq}    {\end{equation}}

\newc{\beqa}    {\begin{eqnarray}}
\newc{\eeqa}    {\end{eqnarray}}
\newc{\bs}    {\section}
\newc{\no}    {\\ \nonumber}

\def\apj{{\em Astrophys. J.  }}
\def\apjl{{\em Astrophys. J. Lett. }}
\def\mnras{{ Mon. Not. Roy. Astron. Soc.  }}
\def\physrep{{\em Phys. Rep.  }}
\def\jcap{{JCAP }}

\def\bx{{\bf{x}}}

\newcommand{\bea}{\begin{eqnarray}}
\newcommand{\eea}{\end{eqnarray}}

\newc{\st}    {\stackrel}

\begin{document}



\title{ Quantum Scales of Galaxies from Self-interacting Ultralight Dark Matter}

\author{Jae-Weon Lee}
\email{scikid@jwu.ac.kr}
\affiliation{Department of Electrical and Electronic Engineering, Jungwon University, 85 Munmu-ro, Goesan-eup, Goesan-gun, Chungcheongbuk-do, 28024, Korea.}

\author{Chueng-Ryong Ji}\email{crji@ncsu.edu}
\affiliation{Department of Physics, North Carolina State University, Raleigh, North Carolina 27695-8202, USA.}

\begin{abstract}
We derive the characteristic scales for physical quantities of dwarf galaxies, such as mass, size, acceleration, and angular momentum, within the self-interacting ultralight dark matter (ULDM) model. Due to the small mass of ULDM, even minor self-interactions can drastically alter these scales
in the Thomas-Fermi limit. 
We suggest that these characteristic scales are connected to  mysteries of observed galaxies.
Oscillation of ULDM field 
can explain the current cosmological density  of  dark matter. Many cosmological constraints suggest that the energy scale $\tilde{m}$ for self-interacting ULDM is typically of the order $10~eV$, whereas the mass $m$ for the non-interacting  case is around $10^{-21}~eV$.
Self-interacting ULDM provides the better explanation for cosmological observations than the non-interacting case.

\end{abstract}

\maketitle

\section{Introduction}
The ultralight dark matter (ULDM) model has emerged as a compelling alternative to cold dark matter (CDM), in which dark matter particles have an exceptionally small mass $m$, typically on the order of $10^{-22}$ eV, and exist in a Bose-Einstein condensate (BEC) state. (For  reviews, see \cite{2009JKPS...54.2622L,2014ASSP...38..107S,2014MPLA...2930002R,2011PhRvD..84d3531C,2016PhR...643....1M,Hui:2016ltb}). This model is known by various other names, including fuzzy DM, BEC DM, scalar field DM, ultra-light axion, and wave-$\psi$ DM~\cite{Baldeschi:1983mq,Sin:1992bg,Lee:1995af,1993ApJ...416L..71W,2000PhRvL..84.3037N,Arbey:2001qi,Hu:2000ke,Peebles:2000yy,2009PhLB..671..174M,2000PhRvD..62j3517S,Alcubierre:2001ea,2012PhRvD..86h3535P,2009PhRvL.103k1301S,Fuchs:2004xe,Matos:2001ps,2001PhRvD..63l5016N,Boehmer:2007um,Eby:2015hsq}.
The long de Broglie wave length $\lambda_{dB}=\hbar/mv\sim kpc$
of ULDM determines the typical size of galaxies, where
$v$ is the typical velocity of the halo dark matter. This scale can give the size and mass  of the smallest galaxies  ~\cite{Lee:2008ux,2009Natur.460..717V} and resolve the small scale issues of CDM including the core-cusp problem, the satellite galaxy plane problem and the missing satellite problem
~\cite{2022JCAP...12..033P,Salucci:2002nc,1996ApJ...462..563N,2003MNRAS.340..657D,2003IJMPD..12.1157T}.
ULDM has also been proposed to address the mysteries of black holes, including the M-sigma relation~\cite{Lee:2015yws} and the final parsec problem~\cite{koo2024final,Bromley:2023yfi}.
In Ref. [\citealp{Lee:2023krm}], the characteristic scales of physical properties
of galaxies, such as angular momentum and mass, in the fuzzy DM model (i.e., non-interacting ULDM) were studied and found to be consistent with observations. 
In this model, quantum pressure arising from the uncertainty principle counteracts the gravitational force.

In the ULDM model,  
dark matter halos of galaxies consist of central cores (solitons) and soliton-like granules surrounding them.  
Note that, in this work, the typical scales apply to these cores and granules
of large galaxies
or to cores of dwarf galaxies, but  not to  galaxies
as a whole. 
These characteristic scales are useful for studying galaxy evolution,  
particularly in dwarf galaxies, which are strongly dominated by dark matter.
One can use these scales to make order-of-magnitude estimates that illuminate the fundamental physics of galaxies.
We emphasize that this model addresses the non-baryonic component of dark matter. Observational data, such as those from the Planck Collaboration, indicate that the total dark matter content in the Universe includes both baryonic and non-baryonic components~\cite{Arbey_2021}. Our model does not aim to account for the baryonic dark matter contribution.

 Fuzzy dark matter (FDM), however, encounters observational challenges, most notably from the Lyman-alpha forest data~\cite{Irsic:2017yje,Armengaud:2017nkf,Lee:2025qvm}, which implies
$m\gtrsim 10^{-21}eV$
~\cite{Zimmermann:2024xvd}. Incorporating self-interactions into the ULDM framework~\cite{Lee:1995af,Boehmer:2007um,Chavanis_2011} has been suggested as a way to alleviate these tensions~\cite{Dave:2023wjq}, accommodating a wider mass range that aligns with cosmological observations~\cite{Hartman:2021upg,Shapiro:2021hjp}. More recently, self-interacting ULDM has also been proposed as a possible solution to the Hubble tension~\cite{Lee:2025roi}, neutrino mass and the electroweak scale problems~\cite{Lee:2024rdc}.

In this paper, we derive typical scales for physical quantities of galaxies in 
self-interacting ULDM  model,
which can be related to mysteries of galaxies.
We compare the characteristic scales predicted by the two models.
In {Section II,} we review the Jeans length of generic ULDM.
In {Section III,} the characteristic scales are derived for fuzzy DM.
In {Section IV,} we consider the self-interacting case. In {Section V,} we discuss the results and outlook.

\section{Jeans length of Ultralight dark matter}

In this section, we review the basics of ULDM.
The ULDM field can be a
   scalar field $\phi$
with an action
\beq
\label{action}
 S=\int \sqrt{-g} d^4x[\frac{-R}{16\pi G}
-\frac{g^{\mu\nu}} {2} \phi^*_{;\mu}\phi_{;\nu}
 -U(\phi)],
\eeq
where the potential for the field 
can be
\beq
U(\phi)=\frac{m^2 c^2}{2\hbar^2}|\phi|^2 + \frac{\lambda |\phi|^4}{4\hbar c}
= \frac{m^2 c^2}{2\hbar^2}|\phi|^2 
+\frac{2 \pi a_s m}{\hbar^2}|\phi|^4.
\label{U}
\eeq
Here, $\lambda={8\pi a_s mc}/{\hbar}$ is a dimensionless coupling
and
$
a_s= \lambda \hbar/(8\pi mc)
$ is a scattering length
~\cite{2011PhRvD..84d3531C}.  FDM corresponds to the case $\lambda = 0$.
In the Newtonian limit, odd-power terms can be ignored because they average out to zero over galactic time scales as the field rapidly oscillates with a frequency of $O(m)$. We adopt the quartic term
with $\lambda>0$, which is the highest even power term that remains renormalizable.
 We 
do not consider the {extension to} cosine potential,
which gives an effective attractive quartic term~\cite{Khlopov:1985fch}.
The evolution of the field is described by the following equation
\beq
\square \phi+2 \frac{d U}{d|\phi|^2} \phi=0,
\eeq
where $\square$ is the d'Alembertian. 
Since galaxies are non-relativistic, 
it is useful to define $\psi$ as
\beq
\phi(t, \mathbf{x})=\frac{1}{\sqrt{2 m}}\left[e^{-i m t} \psi(t, \mathbf{x})+e^{i m t} \psi^*(t, \mathbf{x})\right].
\eeq
 Then the field amplitude becomes~$
|\phi|^2=\frac{\hbar^2}{m^2}|\psi|^2.
$

 In the Newtonian limit,
the macroscopic wave function $\psi$ satisfies the following  nonlinear Schr$\ddot{o}$dinger-Poisson equation (SPE);
\beqa
\label{spe}
i\hbar \partial_{{t}} {\psi} &=&-\frac{\hbar^2}{2m} \nabla^2 {\psi} +m{V} {\psi}+\frac{\lambda \hbar^3}{2cm^3}|\psi|^2\psi, \no
\nabla^2 {V} &=&{4\pi G} \rho,
\eeqa
where the 
 the DM mass density $\rho=m|\psi|^2$, and $V$ is the gravitational potential. 

 Cosmological structure formation
 is described by an equation for the
  density contrast $\delta\equiv
  \delta \rho/\bar{\rho}
  =(\rho-\bar{\rho})/\bar{\rho}
  =\sum_k \delta_k e^{ik\cdot r}$ with a wave vector $k$ of the perturbation~\cite{Chavanis_2012},
   \beq
  \frac{d^2 \delta_k}{d t^2} +  \left[(c^2_q+c^2_s)k^2-4\pi G \bar{\rho} \right]\delta_k=0,
 \eeq
 where $\bar{\rho}$ is the cosmological  background dark matter density, $c_q=\hbar k/2m$ is a quantum velocity
and   
$c_s=\sqrt{{4 \pi a_s \hbar^2 \bar{\rho}}/{m^3}}
={\sqrt{{{\hbar}^3 \lambda  \bar{\rho} }/{2 c m^4}}}
$ is the sound velocity 
from self-interaction.
The Jeans length 
corresponds to the
wave vector $k_J$ satisfying
$(c^2_q+c^2_s)k^2-4\pi G \bar{\rho} = 0$.
The DM system becomes unstable to perturbations with $k < k_J$, resulting in the formation of cosmic structures.

During the structure formation, gravity dominates over pressure, and we have the equation:
$ \frac{d^2 \delta_k}{dt^2} \simeq 4\pi G \bar{\rho} \delta_k.$
This leads to a typical time scale for structure formation~\cite{2011PhRvD..84d3531C}:
\beq
\label{tc}
t_c = \frac{1}{\sqrt{G \bar{\rho}}}, 
\eeq
which is approximately the order of the Hubble time at the time of structure formation. 
Therefore, $t_c$ depends on the redshift at which individual galaxies form. 
However, this does not mean that all galaxies have the same time scale. In fact, $t_c$ represents the time scale of the lightest galaxy or cores of heavy dark matter halos formed at a given redshift. 
This time scale can be further justified by the following argument. An astronomical object like
a dwarf galaxy with a typical mass $M$ and radius $R$ has a characteristic time scale given by:
$\frac{R}{\sqrt{GM/R}} \simeq \frac{1}{\sqrt{G \rho_c}}, $
where $\rho_c = O(10^2) \bar{\rho}$ is the typical density of the object. Therefore, Eq. (\ref{tc}) provides a reasonable estimate for the characteristic time scale in galactic dynamics.

\section{Quantum Scales of fuzzy dark matter}
In this section we 
review the case of  the free field model ($\lambda=0$), i.e., FDM model~\cite{Lee:2023krm}.
From the condition
$c^2_q k^2=4\pi G \bar{\rho}$,
   one can obtain
   the quantum Jeans length scale~\cite{Hui:2016ltb,Khlopov:1985fch} at a redshift $z$:
\beq
\label{lambdaQ}
\lambda_Q(z)= \frac{2\pi}{k}=\left(\frac{\pi^3 \hbar^2 }{Gm^2\bar\rho(z)}\right)^{1/4}
= 71.75~kpc
\left(\frac{{m}}{10^{-22}~eV}\right)^{-1/2}\left(\frac{\bar{\rho}}{10^{-7}M_\odot/pc^3}\right)^{-1/4}
\propto (1+z)^{-3/4},
\eeq
where $m_{22}=m/10^{-22}eV$.

A different length scale, characterizing a stable galactic core, emerges from the equilibrium between the self-gravitational force of a ULDM soliton with a mass $M$ and quantum pressure;
\beq
\label{dB}
R_{99} =9.95\left(\frac{\hbar}{m}\right)^2 \frac{1}{ G M}= 8.5~kpc\left(\frac{10^{-22}eV}{m}\right)^2 \frac{10^8 M_\odot}{M},
\eeq
where $R_{99}$ is the radius containing 99\% of the ULDM mass
~\cite{PhysRev.187.1767}.
We choose $R_{99}$ as a length scale $\bx_c$ for FDM, which is of the order of the
de Broglie wavelength of the ULDM particles in the soliton.
$R_{99}$ is inversely proportional to the soliton mass, which is not
typical of observed galactic cores.

The scales of ULDM systems are determined by two mass parameters $m$ and $M$.
From $\lambda_Q$ the quantum Jeans mass can be derived as
\beq
\label{MQ}
M_{Q}(z)
=\frac{4\pi}{3} \left(\frac{\lambda_Q}{2}\right)^3\bar{\rho}=\frac{ \pi^{\frac{13}{4}}}{6}
\left(\frac{\hbar}{G^{\frac{1}{2}}   m}\right)^{\frac{3}{2}} \bar{\rho}(z)^\frac{1}{4}
 =
1.93\times 10^7~M_\odot\left(\frac{{m}}{10^{-22}~eV}\right)^{-3/2}\left(\frac{\bar{\rho}}{10^{-7}M_\odot/pc^3}\right)^{1/4}
 \propto (1+z)^{3/4},
\eeq
which is the typical mass
scale of dwarf galaxies or galactic cores and appropriate 
for $M$.
This does not imply that all galaxies have identical   masses.  
Simulations suggest that the core mass scales as $M_h^{1/3}$, where $M_h$ is the total halo mass~\cite{Schive:2014hza}.

Inserting $M=M_Q$ into Eq. (\ref{dB}) gives
\beq
\label{R99}
\bx_c=R_{99}=
\frac{1.44\sqrt{\frac{\hbar}{m}}}{ (G \bar{\rho} )^{1/4}}
=
43.97~kpc\left(\frac{{m}}{10^{-22}~eV}\right)^{-1/2}\left(\frac{\bar{\rho}}{10^{-7}M_\odot/pc^3}\right)^{-1/4},
\eeq
which is $0.61~\lambda_Q$.
$R_{99}$ is proportional to $\lambda_Q$ but slightly smaller than $\lambda_Q$. (See Fig. 1)

From these fundamental typical scales
and Eq. (\ref{MQ}) one can easily obtain other physical scales.
The typical acceleration scale is given by
\beq
a_c=\bx_c/t_c^2 =0.0044 ~G^3 m^4 M^3/\hbar^
4 =8.38 \times 10^{-14}~meter/s^2\left(\frac{m}{10^{-22}eV}\right)^4\left(\frac{M}{10^8 M_\odot}\right)^3 
\simeq  \sqrt{\frac{\hbar}{m}} (G \bar{\rho})^{3/4},
\label{ac}
\eeq
which is similar to the Modified Newtonian dynamics (MOND) scale
$a_0=1.2\times 10^{-10}~meter/s^2$
for $m\simeq 10^{-21}~eV$.
In Ref. [\citealp{Lee:2019ums}], it is suggested that MOND arises as an effective phenomenon of ULDM, with $a_c$ linked to  the observed radial acceleration relation,
if galactic cores have a similar mass $M$.
In that scenario 
ULDM halos have a solitonic core and 
granules
interacting with
baryons. 
At a specific radius, the total density of dark matter and baryons follows an isothermal profile, where the acceleration matches $a_c$, possibly explaining the empirical radial acceleration relation. 
However, precise numerical analysis is needed to validate this scenario.

Using $\bar{\rho}=(1296 G^3 m^6 M^4)/(\hbar^6 \pi^{13})$ from Eq. (\ref{MQ})  one can obtain
the typical velocity  
\beq
v_c\equiv \bx_c/t_c= 0.21~G M m/\hbar
=4.71~km/s \left(\frac{M}{10^8 M_\odot}\right) \left(\frac{m}{10^{-22}eV}\right)\simeq  \sqrt{\frac{\hbar}{m}} (G \bar{\rho} )^{1/4},
\eeq
which corresponds to a typical velocity in dwarf galaxies
and
leads to a typical angular momentum of galactic halos
\beq
L_c=M \bx_c v_c =2.09~{\hbar}\frac{ M}{m}
=2.33\times 10^{96}~\hbar ~\left(\frac{M}{10^8 M_\odot}\right) \left(\frac{10^{-22}eV}{m}\right)
\simeq \frac{\left(\frac{\hbar}{m}\right)^{5/2}\bar{\rho} ^{1/4}}{G^{3/4}}.
\eeq
This implies that the typical angular momentum scales with the number of particles times  $\hbar$, which clearly reveals the quantum nature of the model.  
The typical density   is
\beq
\rho_c
\equiv M/\bx_c^3=0.001\frac{G^3 m^6 M^4}{ {\hbar}^6}
=1.62\times 10^{-4}~M_\odot/pc^3~\left(\frac{m}{10^{-22}eV}\right)^6\left(\frac{M}{10^8 M_\odot}\right)^{4}
\propto (1+z)^3,
\eeq
which is
independent of $m$ and implies  
a high density at a high redshift.  
This could facilitate the early formation of supermassive black holes~\cite{Chiu:2025vng}.

The typical surface density $\Sigma\equiv M/x_c^2 $
is
\beq
\Sigma
=
0.01\frac{G^2 m^4 M^3 }{ \hbar^4}
=
1.38~M_\odot/pc^2\left(\frac{{m}}{10^{-22}eV}\right)^{4} \left(\frac{M}{10^8 M_\odot}\right)^4 \propto (1+z)^3
\eeq
which fails to explain the observed data~\cite{Burkert:2020laq,DeLaurentis:2022nrv}.
Since Eq.~(\ref{dB}) predicts that smaller cores are more massive and exhibit higher values of $\Sigma$, this contradicts observations indicating that $\Sigma$ remains approximately constant at $ 75 M_\odot/{\rm pc}^2$.

The scale for the gravitational potential,
\beq
V_c \  =
\frac{GM}{\bx_c}
=\frac{0.1~m^2}{\hbar^2} \left({ G M}\right)^{2}=5.6\times 10^{-10}c^2~\left(\frac{m}{10^{-22}eV}\right)^2\left(\frac{M}{10^8 M_\odot}\right)^{2} ,
\eeq
becomes relativistic when $M\simeq 10^{12} M_\odot$ for $m=10^{-21}eV$.
This mass $M$ is similar to the maximum galaxy mass.
All of the above constraints on galactic scales appear to favor a mass of approximately $m \simeq 10^{-21}~\text{eV}$.

   \section{Self-interacting case}
In this section we consider the self-interacting case with a quartic potential 
$U(\phi)$ in Eq. (\ref{U})
with $\lambda>0$ ~\cite{Lee:1995af}.   
In the Thomas-Fermi (TF) limit, the kinetic term can be ignored and the physical quantities often depend on the single parameter
$\tilde{m}\equiv m/\lambda^{1/4}$, which
represents the typical energy scale of $\phi$.

   In this limit the equation for the
  density contrast $\delta$
  is now
   \beq
  \frac{d^2 \delta_k}{d t^2} +  \left[(c^2_s)k^2-4\pi G \bar{\rho} \right]\delta_k=0.
 \eeq

From the equation one can obtain
the Jeans length from
self-interaction~\cite{PhysRevD.42.3918},
\beq
\label{lambdaJ}
\lambda_J =  2 \pi {\hbar}  \sqrt{\frac{a_s}{G m^3}}
=2 R_{TF}
={\sqrt{\frac{\pi\hbar^3 \lambda }{2c G m^4}}}
= 0.978~kpc\left(\frac{\tilde{m}}{10eV}\right)^{-2},
\eeq 
which does not depend on the density
and time, unlike  $\lambda_Q$ of FDM.

In the TF limit, the exact ground state solution of SPE is given by ~\cite{Lee:1995af,2011PhRvD..84d3531C}
\beq
|\psi|^2=\frac{|\psi(0)|^2 R_{TF}}{\pi r} sin\left(\frac{\pi r}{R_{TF}}\right),
\eeq
where
the soliton size is 
\beq
R_{TF}=\pi \hbar \sqrt{\frac{a_s}{Gm^3}}
={\sqrt{\frac{\pi\hbar^3 \lambda }{8c G m^4}}}.
\eeq

From $\lambda_J$ we can
obtain the Jeans mass
for ULDM,
\beq
\label{MJ}
M_J(z)\equiv\frac{4\pi}{3} \left(\frac{\lambda_J}{2}\right)^3\bar{\rho}=\frac{4\pi^4 \hbar^3}{3} \left(\frac{ a_s }{G m^3}\right)^{3 / 2} \bar{\rho}
=\frac{\pi^{5/2}}{\sqrt{288}}\left(\frac{\hbar^3 \lambda }{c G m^4}\right)^{3/2} \bar{\rho}
= 49 ~M_\odot\left(\frac{\tilde{m}}{10eV}\right)^{-6}\left(\frac{\bar{\rho}}{10^{-7}M_\odot/pc^3}\right) \propto 
 (1+z)^3,
\eeq
where $z$ is the redshift when the perturbation starts to grow, and
$\bar{\rho}(z)=\rho(0)(1+z)^3$.
Since the present value of $\bar{\rho}$
is about $ 10^{-7} M_\odot/pc^3$,
$M_J(z \simeq 0)$ is too small to explain
the observed galaxies,
if they are formed
at low redshifts.
This discrepancy can be resolved by assuming that the collapse of density perturbations leading to galaxy formation began at very high redshifts ($z \gg 10$), as recent James Webb observations suggest~\cite{Yan_2022,Carniani_2024}.
To account for the early formation of high-redshift galaxies, dark matter perturbations must begin growing at significantly higher redshifts.
For example, $M_J(z=100)\simeq 5\times 10^7 M_\odot$,
which is similar to the  mass of the smallest galaxies.
Somewhat smaller values of $\tilde{m}$ can also help increase $M_J$.

The TF limit corresponds to  
$\lambda_J > \lambda_Q$, which
implies ~\cite{Chavanis_2011}
\beq
\label{lambda}
\lambda 
 > 5.3\times 10^{-89}
 \left(\frac{m}{10^{-22} eV}\right)^{3} 
\left(\frac{\bar{\rho}(z)}{10^{-7} M_\odot/pc^3}\right)^{-1/2}.
\eeq
This corresponds
to the condition that
the spatial size of
initial dark matter density perturbations
is determined by the repulsion not by the quantum pressure.
(See Fig. 2.)
If we also impose the condition $R_{TF} > R_{99}$,  
we obtain the bound  
\beq
\lambda > 3.02 \times 10^{-90} \left( \frac{M}{10^8\, M_\odot} \right)^{-2},
\eeq
which notably does not depend on the particle mass $m$.
This condition ensures that galactic cores are stabilized by the repulsive force arising from the quartic interaction.

\begin{figure}[]
\includegraphics[width=0.7\textwidth]{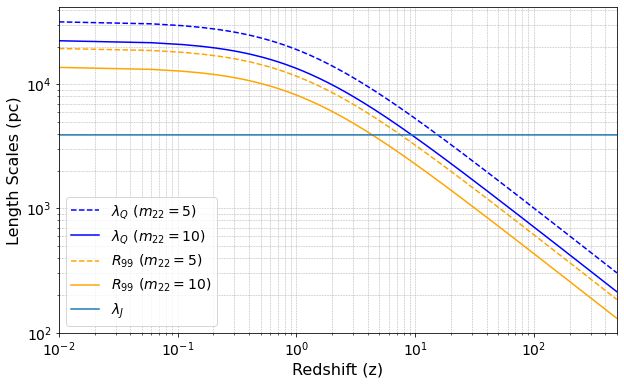}
\caption{ The evolution of the 
typical length scales of galaxies versus  redshift $z$. 
The upper two curves represent $\lambda_Q$
for $m_{22}=m/10^{-22}eV=5$ and  $m_{22}=10$, respectively,
while the lower two curves represent the corresponding $R_{99}$. 
The horizontal line represents $\lambda_J=2R_{TF}$ in the TF limit
with $\tilde{m}=5~eV$.
When $\lambda_J > \lambda_Q$, the size and mass of a dark matter halo formed at that time are governed by self-interaction-induced repulsion rather than quantum pressure, especially at high redshifts $z\gg 10$ and for larger mass
$m\gtrsim 10^{-21}~eV$. We expect galaxy density perturbations to begin collapsing in the redshift range $O(10) \lesssim z \lesssim O(100)$.}
\label{length}
\end{figure}

For $\lambda_J>\lambda_Q$, it is natural to represent
physical quantities with 
$M=M_J$ and $\tilde{m}$.
The dynamical time scale $t_c \sim \frac{1}{\sqrt{G \bar{\rho}}}$ becomes~\cite{2011PhRvD..84d3531C}
\beq
t_c=\left(\frac{R_{TF}^3}{G M}\right)^{1 / 2}
=\pi ^{3/4}   \left(\frac{\hbar^9 \lambda ^3}{512c^3 G^5m^{12}}\right)^{1/4}  \sqrt{\frac{1}{M}}
\propto (1+z)^{-3/2}.
\eeq
We have the length scale
$x_c=R_{TF}$, the time scale $t_c$,
and the mass scale $M=M_J$ for the self-interacting ULDM, from which
we can derive the typical scales
for various other physical quantities as follows.
Since  $x_c$, $t_c$ and $M$
are functions of $\lambda/m^4$, we expect many derived quantities from them to have a dependency on $\tilde{m}\equiv m/\lambda^{1/4}$.

Then, the typical density  $\rho_c\equiv M/\bx_c^3$ is
\beq
\rho_c=
\frac{2\sqrt{2} M \left(\frac{c G m^4}{\lambda }\right)^{3/2}}{\hbar^{9/2} \pi ^{3/2}}
=0.106~M_\odot/pc^3~\left(\frac{\tilde{m}}{10~eV}\right)^6\left(\frac{M}{10^8 M_\odot}\right) \propto (1+z)^3,
\eeq
which is independent of $\tilde{m}$.

The typical velocity is 
\beq
v_c\equiv \bx_c/t_c
=\frac{2^{7/4}\left(\frac{c G^3 m^4}{\hbar^3 \lambda }\right)^{1/4}}{ \pi ^{1/4}  } \sqrt{{M}}
=59.28~km/s \left(\frac{M}{10^8 M_\odot}\right)^{1/2} \left(\frac{\tilde{m}}{10 eV}\right)\propto (1+z)^{3/2},
\eeq
which is similar to the typical velocity dispersion in a dwarf galaxy.
$v_c$ leads to a typical angular momentum in turn
\beq
L_c=
M \bx_c v_c 
= \left(\frac{32 \pi G \hbar^3 \lambda }{cm^4}\right)^{1/4}{M}^{3/2}
=3.375\times 10^{96}~\hbar ~\left(\frac{M}{10^8 M_\odot}\right)^{3/2} \left(\frac{10~eV}{\tilde{m}}\right)
\propto (1+z)^{9/2},
\eeq
which is not
of  the order of
$\hbar(M/m)$ 
as in the fuzzy dark matter.

The typical acceleration scale is given by
\beq
a_c
=\bx_c/t_c^2 
=\frac{16 c G^2 m^4M}{ \pi   \hbar^3 \lambda }
=1.163\times 10^{-10}~meter/s^2\left(\frac{\tilde{m}}{10 eV}\right)^4\left(\frac{M}{10^8 M_\odot}\right) \propto (1+z)^3
\label{ac}
\eeq
which is  similar
to the  MOND scale
$a_0=1.2\times 10^{-10}~meter/s^2$
for $M=10^8M_\odot$.
 Galaxies with similar core mass $M$ are expected to exhibit similar values of $a_c$.

The scale for the gravitational potential,
\beq
V_c  
= G M/\bx_c
=\frac{2 a_s \hbar^2 \pi^3 \bar{\rho}}{3 m^3}
=\frac{G M \sqrt{\frac{2}{\pi}}}{\sqrt{\frac{\hbar^3 \lambda}{c G m^4}}}
=4.888\times 10^{-9}c^2~\left(\frac{\tilde{m}}{10~eV}\right)^2\left(\frac{M}{10^8 M_\odot}\right) \propto (1+z)^3,
\eeq
 becomes relativistic when $M\simeq 10^{16} M_\odot$ for $\tilde{m}=10~eV$.

The typical surface density $\Sigma$
is
\beq
\Sigma=\frac{\hbar \pi^{3/2} \sqrt{\frac{\hbar \lambda}{c G m^4}} \bar{\rho}}{6 \sqrt{2}}
= 5.124 ~M_\odot/pc^2\left(\frac{\tilde{m}}{10eV}\right)^{-2}\left(\frac{\bar{\rho}}{10^{-2}M_\odot/pc^3}\right)
=104.4 ~M_\odot/pc^2\left(\frac{\tilde{m}}{10eV}\right)^{-2} \left(\frac{M}{10^8 M_\odot}\right) \propto (1+z)^3
\eeq
which is  similar to the observed
value $\Sigma \simeq 75M_\odot/pc^2$
for $M=10^8 M_\odot$.
Of course, $\Sigma$
in the TF limit is also redshift dependent. However, if the dwarf galaxies
observed start forming around the same redshift, for example, $z \simeq 100$, they can exhibit a comparable  $\Sigma$.
The dependence of $\Sigma$ on $M$ is weaker than in the FDM case.

Considering all these factors, self-interacting ULDM offers a more compelling explanation for the observed galaxies if their cores formed at similarly high redshifts.
To be conclusive, we need a high-resolution simulation of structure formation with self-interacting ULDM, as cores in a massive halo may exhibit slightly  different properties than those of a single soliton.

\begin{figure}[]
\includegraphics[width=0.7\textwidth]{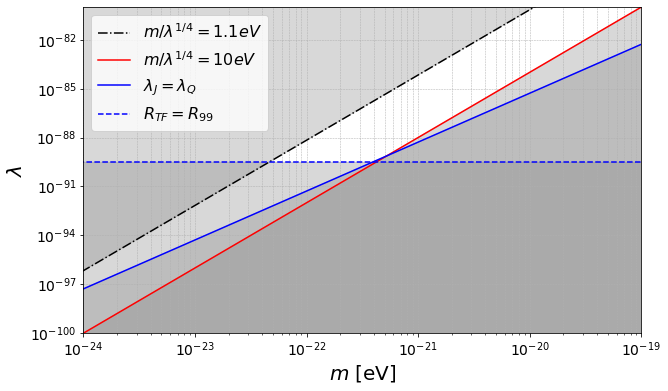}
\caption{ 
Cosmological constraints on the mass $m$ and coupling constant $\lambda$ of ULDM are shown. The gray region denotes the parameter space excluded by current constraints. The dash-dot line corresponds to the bound from the dark matter density $\Omega_\phi$ (Eq. (\ref{density})). The red solid line indicates the observational constraint from galaxies, $\tilde{m} \lesssim 10~\text{eV}$. The blue line marks the boundary where the condition $\lambda_J > \lambda_Q$ (Eq. (\ref{lambda})) holds at redshift $z = 100$. The dashed horizontal line represents the condition $R_{TF} > R_{99}$ for a soliton mass of $M = 10^8~M_\odot$.
}
\label{omegafig}
\end{figure}

To derive the dark matter density, we follow the approach of Ref.[\citealp{Hui:2016ltb}].
For $\lambda=0$, the field begins to oscillate at $\phi = F$ when the Hubble parameter satisfies $H \sim \frac{T_{\text{osc}}^2}{m_P} = m$,
where $m_P = 1/\sqrt{8\pi G}$ is the reduced Planck mass, and $T_{osc}=\sqrt{mm_P}$ represents the temperature at that time.
The typical energy
density of ULDM at this time is of order of 
$F^2 m^2$.
At the matter-radiation equality with the temperature $T_{eq}\simeq 1 eV$ 
we find the relation
$
\frac{{m}^2 F^2}{T_{osc}^4} \frac{T_{osc}}{T_{eq}} \sim 1$.
Therefore,
\beq
F \sim \frac{m_{{P}}^{3 / 4} T_{eq}^{1 / 2}}{m^{1 / 4}} \sim  10^{17} \mathrm{GeV}
\left( \frac{10^{-22}~eV}{m}\right)^{1/4},
\eeq
which indicates that the typical value of $\phi$ lies near the Grand Unified Theory (GUT) scale.
From the present Hubble parameter $H_0$ and the temperature of the universe one can estimate the  density parameter today for ULDM \cite{Hui:2016ltb}, 
\beq
\Omega_{\phi} \sim 0.1\left(\frac{F}{10^{17}\mathrm{GeV}}\right)^2\left(\frac{m}{10^{-22} \mathrm{eV}}\right)^{1 / 2}. 
\eeq

For $\lambda \neq 0$, the above logic used for the $\lambda = 0$ case cannot be directly applied, as the equation of state for $\phi$ depends on the field value~\cite{Li:2013nal}.
To treat the oscillation of $\phi$ as cold dark matter, the quartic term in $U(\phi)$ must be smaller than the quadratic term, at least at the time when the temperature is $T_{eq}$. 
 That is, $\phi_{\text{eq}} < m/\sqrt{\lambda}$ must hold at that time.
Considering oscillation
of the field one can obtain
the present density parameter
\beq
\Omega_\phi \simeq \frac{{\frac{m^2\phi_{eq}^2}{2}}\left(\frac{T_{\text{now}}}{T_{\text{eq}}}\right)^3}{3H_0^2 m^2_P}
\lesssim \frac{m^4}{6\lambda  H_0^2 m_P^2} \left(\frac{T_{\text{now}}}{T_{\text{eq}}}\right)^3
\eeq
which leads to
\beq
\phi_{\text{eq}} = \sqrt{\frac{6 \Omega_\phi H_0^2 m_P^2}{m^2} \left(\frac{T_{\text{eq}}}{T_{\text{now}}}\right)^3} \simeq 1.2 \times 10^{13}~\text{GeV} 
\left(\frac{10^{-22~\text{eV} }}{m}\right),
\eeq
and ~\cite{Boudon:2022dxi} 
\beq
\label{density}
\tilde{m} = m/\lambda^{1/4} \gtrsim \left({6 \Omega_\phi} \left(\frac{T_{\text{eq}}}{T_{\text{now}}}\right)^3 H_0^2 m_P^2 \right)^{1/4} \simeq 1.1~\text{eV},
\eeq
where $T_{now}=2.3\times 10^{-4}~eV$
is the current temperature
of the universe and $\Omega_\phi\simeq 0.26.$

On the other hand, the cross-section inferred from galaxy cluster collisions suggests $a_s^2/m \lesssim 1~cm^2/g$.
\cite{Hartman:2021upg,Garcia:2023abs,Mishchenko:2015ysp},
which leads to
\beq
\lambda < 5.3 \times 10^{-44} \left(\frac{m}{10^{-22}~eV} \right)^{3/2}.
\eeq
Taking all these into account, we expect
$1~eV\lesssim \tilde{m}
\lesssim 10~eV$
for $m\lesssim 10^{-5}~eV$.
One can also obtain the field value at the onset of oscillations:
\beq
\label{phiosc2}
\phi_{osc}=\frac{{\pi}^{4/3} {m_P} \left(\frac{{g_*} {T_{eq}}}{{\tilde{m}}}\right)^{2/3}}{30^{2/3}}
= 4.8\times 10^{17}GeV
\left(\frac{\tilde{m}}{10~eV}\right)^{-2/3},
\eeq
where $g_* = 3.36$ is the number of relativistic degrees of freedom.
(See Ref. [\citealp{Lee:2024rdc}] for details.)

Fig. \ref{omegafig} illustrates the allowed parameter regions that satisfy these constraints.
 From the figure, cosmological constraints appear to favor the fiducial values $(m \sim 10^{-21}~\text{eV}, \lambda \sim 10^{-88})$, if we also incorporate constraints from the effective number of relativistic species during Big Bang nucleosynthesis
($10^{-19}~eV\lesssim m\lesssim 10^{-21}~eV$) ~\cite{Li:2013nal}.

\section{Results and Outlook}

Through the analysis of the SPE derived from $\phi^4$ scalar field theory, we have estimated the characteristic scales of physical quantities observed in galaxies for both the free $(\lambda=0)$ and self-interacting $(\lambda>0)$ ULDM scenarios.
 By accounting for both quantum pressure and self-interaction pressure, the limitations of each model can be addressed. 
 For parameters consistent with observations, 
 dark matter halos associated with galaxies are expected to begin collapsing at high redshifts ($z \gtrsim 100$) in the TF limit.
 Compared to $\lambda$, the mass $m$ of self-interacting ULDM is subject to fewer known constraints, warranting further investigation.

 The inferred density and typical field values of ULDM suggest a possible connection to GUT-scale physics.
 The existence of ULDM oscillations with a frequency $\sim m$  could be detectable through pulsar timing array experiments ~\cite{NANOGrav:2023hfp}, or atomic clock experiments
~\cite{Kouvaris:2019nzd}.
Efforts to detect ULDM oscillations via precision frequency measurements rely on the premise that ULDM induces minute fluctuations in fundamental constants
which, in turn, can be detected as variations in the frequencies of atomic clocks.
The gravitational lensing 
~\cite{Boudon:2023vzl} and the gravitational-wave signatures ~\cite{GalazoGarcia:2025xjh} can also offer a promising avenue to detect or constrain self-interacting ultralight dark matter. The characteristic scales derived in this work offer a convenient means of estimating key physical quantities of galaxies.

 Future observations
 of early galaxies, including data from the James Webb Space Telescope~\cite{Labb__2023}, will provide new insights into galaxy evolution, potentially validating these characteristic scales.
 Exploring characteristic scales arising from ultralight axions with a cosine potential constitutes a promising direction for future research.\\

\subsection*{Acknowledgments}
This work was supported in part by the U.S. Department of Energy (Grant No. DE-FG02-03ER41260).
This research used resources of the National Energy Research
Scientific Computing Center, a DOE Office of Science User Facility
supported by the Office of Science of the U.S. Department of Energy
under Contract No. DE-AC02-05CH11231 using NERSC award
NP-ERCAP0027381.
CRJ thanks for the hospitality during his visit to the Asia Pacific Center for Theoretical Physics (APCTP) 
where this work was completed.


\end{document}